# Mechanics of granular column collapse in fluid at varying slope angles

K. Kumar
Computational Geomechanics Research Group, Department of Engineering
University of Cambridge, UK

J-Y. Delenne
IATE, UMR 1208 INRA-CIRAD-Montpellier Supagro-UM2,
University of Montpellier 2, France.

K. Soga*
Department of Civil and Environmental Engineering,
University of California, Berkeley, USA
soga@berkeley.edu

# Abstract

This paper investigates the effect of initial volume fraction on the runout characteristics of collapse of granular columns on slopes in fluid. Two-dimensional sub-grain scale numerical simulations are performed to understand the flow dynamics of granular collapse in fluid. The Discrete Element (DEM) technique is coupled with the Lattice Boltzmann Method (LBM), for fluid-grain interactions, to understand the evolution of submerged granular flows. The fluid phase is simulated using Multiple-Relaxation-Time LBM (LBM-MRT) for numerical stability. In order to simulate interconnected pore space in 2D, a reduction in the radius of the grains (hydrodynamic radius) is assumed during LBM computations. The collapse of granular column in fluid is compared with the dry cases to understand the effect of fluid on the runout behaviour. A parametric analysis is performed to assess the influence of the granular characteristics (initial packing) on the evolution of flow and run-out distances for slope angles of 0°, 2.5°, 5° and 7.5°. The granular flow dynamics is investigated by analysing the effect of hydroplaning, water entrainment and viscous drag on the granular mass. The mechanism of energy dissipation, shape of the flow front, water entrainment and evolution of packing density is used to explain the difference in the flow characteristics of loose and dense granular column collapse in fluid.

*Keywords*: Lattice Boltzmann (LBM), Discrete Element Method (DEM), Granular column collapse, Granular flows, Hydroplaning, Water entrainment, Viscous drag.

# 1. Introduction

Catastrophic earth movement events, such as landslides, debris flows, rock avalanches and reservoir embankment failures, exemplify the potential consequences of an earth gravitational instability. Slope failure is a problem of high practical importance for both civil engineering structures and natural hazard management. The study described in this paper examines the stability of underwater slopes, which are caused by excess seepage or earthquakes. They can damage offshore structures nearby and may generate a tsunami.

In order to describe the mechanism of underwater granular flows, it is necessary to consider both the dynamics of the solid phase of granular matter and the role of the ambient fluid, which exists either inside the pores of the granular body and as free water outside the granular body [1, 2]. Initial acceleration plays a crucial role in underwater landslide propagation [3], as the initial acceleration increases, there is a limited time for the landslide to deform during the acceleration phase. The initiation and propagation of submarine granular flows depend mainly on geometry (e.g. slope angle, lateral extent, etc.), initial stress conditions, density, soil properties, and the quantity of the material destabilised. Although certain macroscopic models are capable of capturing simple mechanical behaviour [4], the complex fundamental mechanism that occurs at the grain scale, such as hydrodynamic instabilities, the formation of clusters, collapse, and transport, require further investigation in order to make better engineering assessment of the potential risk of damages against underwater slope failures for example.

The momentum transfer between the discrete and the continuous phases of fluid saturated granular material significantly affects the dynamics of the flow [5]. The grain-scale description of the granular material enriches the macro-scale variables. In particular, when the solid phase reaches a high-volume fraction, it is important to consider the strong heterogeneity arising from the contact forces between the grains, the drag interactions which counteract the movement of the grains, and the hydrodynamic forces that reduce the weight of the grains inducing a transition from a dense compacted to a dense suspended flow [6].

The case of granular material movements in presence of an interstitial fluid at the grain-scale has been less studied. In this paper, we report the findings of the study on the granular column collapse in fluid in the inclined configuration using the coupled Lattice Boltzmann Method (LBM) and Discrete Element Method (DEM). We examined the effect of density and slope angle on the runout evolution.

# 2. LBM formulation

2.1 General

The Lattice Boltzmann Method (LBM) is a 'micro-particle' based numerical time-stepping procedure for the solution of incompressible fluid flows. Consider a 2D incompressible fluid flow with density ρ and kinematic viscosity $v$, in a rectangular domain D. The fluid domain is divided into a rectangular grid or lattice, with the same spacing 'h' in both the x- and the y-directions, as shown in Figure 1. The present study focuses on two-dimensional problems, hence the D2Q9 momentum discretisation is adopted (see He et al., [7] for naming convention).

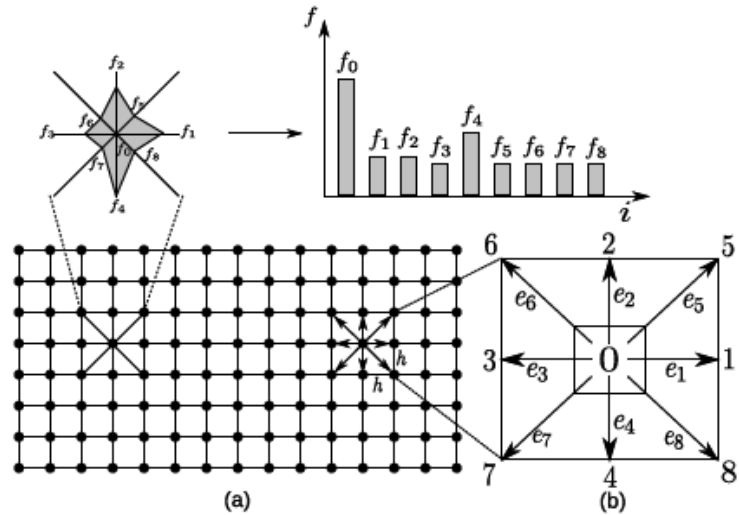

Figure 1: The Lattice Boltzmann discretisation and D2Q9 scheme: (a) a standard LB lattice and histogram views of the discrete single particle distribution function/direction-specific densities; (b) D2Q9 model

The lattice Boltzmann Bhatnagar-Gross-Krook (LBGK) method is capable of simulating various hydrodynamics [8] and offers intrinsic parallelism. Although LBM is successful in modelling complex fluid systems, such as multiphase flows and suspensions in fluid, LBM may lead to numerical instability when the dimensionless relaxation time 'τ' is close to 0.5. The Multi-Relaxation Time Lattice Boltzmann Method (LBM-MRT) overcomes the deficiencies of linearised single relaxation LBM-BGK, such as fixed Prandtl number (Pr=$v$/κ), where the thermal conductivity 'κ' is unity [9].

The LB-MRT model offers better numerical stability and has more degrees of freedom. In the formulation of the linear Boltzmann equation with multiple relaxation time approximation, the lattice Boltzmann equation is written as:

$$f_\alpha(x + e_i \Delta_t, t + \Delta_t) - f_\alpha(x, t) = -S_{\alpha i} \tag{1}$$

where S is the collision matrix. The nine eigenvalues of **S** are all between 0 and 2 so as to maintain linear stability and the separation of scales, which means that the relaxation times of non-conserved quantities are much faster than the hydrodynamic time scales. The LBGK model is the special case in which the nine relaxation times are all equal and the collision matrix $\mathbf{S} = \frac{1}{\tau}\mathbf{I}$, where **I** is the identity matrix. The evolutionary progress involves two steps; advection and flux. The advection can be mapped to the momentum space by multiplying through by a transformation matrix M and the flux is still finished in the velocity space. The evolutionary equation of the multi-relaxation time lattice Boltzmann equation is written as:

$$\mathbf{f}(\mathbf{x} + e_i \Delta_t, t + \Delta_t) - \mathbf{f}(\mathbf{x}, t) = -M^{-1}\hat{\mathbf{S}}(\hat{\mathbf{f}}(\mathbf{x}, t) - \hat{\mathbf{f}}^{eq}(\mathbf{x}, t)) \tag{2}$$

where **M** is the transformation matrix mapping a vector **f** in the discrete velocity space $V = \mathbb{R}^b$ to a vector $\hat{\mathbf{f}}$ in the moment space $V = \mathbb{R}^b$.

$$\hat{\mathbf{f}} = M\mathbf{f} \tag{3}$$

$$\mathbf{f}(\mathbf{x}, t) = [f_0(\mathbf{x}, t), f_1(\mathbf{x}, t), \ldots f_8(\mathbf{x}, t)]^T \tag{4}$$

The collision matrix $\hat{\mathbf{S}} = MSM^{-1}$ in the moment space is a diagonal matrix: $\hat{\mathbf{S}} = \text{diag}[s_1, s_2, s_3, \ldots s_9]$. The transformation matrix **M** can be constructed via the Gram-Schmidt orthogonalisation procedure. Through the Chapman-Enskog expansion [10], the incompressible Navier-Stokes equation can be recovered and the viscosity is given as:

$$v = c_s^2 \Delta t (\tau - 0.5) \tag{5}$$

## 2.2 Turbulence in Lattice Boltzmann Method

Modelling fluids with low viscosity like water remains a challenge, necessitating very small values of $h$, and/or $\tau$ very close to 0.5 [7]. Turbulent flows are characterised by the occurrence of eddies with multiple scales in space, time and energy. In this study, the Large Eddy Simulation (LES) is adopted to solve turbulent flow problems. The separation of scales is achieved by filtering of the Navier-Stokes equations, from which the resolved scales are directly obtained and unresolved scales are modelled by a one-parameter Smagorinsky sub-grid methodology. It assumes that the Reynold's stress tensor is dependent only on the local strain rate [11]. The turbulent viscosity ν is related to the strain rate $S_{ij}$ and a filtered length scale 'h' as follows:

$$v_t = (S_c h)^2 \overline{S} \quad (6)$$

$$\overline{S} = \sqrt{\sum_{ij} \tilde{S}_{ij} \tilde{S}_{ij}} \quad (7)$$

where $S_c$ is the Smagorinsky constant found to be close to 0.03 [12].

The effect of the unresolved scale motion is considered by introducing an effective collision relaxation time scale $\tau_t$, so that the total relaxation time $\tau_*$ is written as:

$$\tau_* = \tau + \tau_t \quad (8)$$

where $\tau$ and $\tau_*$ are respectively the standard relaxation times corresponding to the true fluid viscosity $v$ and the turbulence viscosity $v_t$, defined by a subgrid turbulence model. The new viscosity $v_*$ corresponding to $\tau_*$ is defined as:

$$v_* = v + v_t = \frac{1}{3}(\tau + \tau_t - 0.5)C^2 \Delta t \quad (9)$$

$$v_t = \frac{1}{3}\tau_t C^2 \Delta t \quad (10)$$

The Smagorinsky model is easy to implement and the Lattice Boltzmann formulation remains unchanged, except for the use of a new turbulence-related viscosity $\tau_*$. The component $s_1$ of the collision matrix $s_1$ becomes $s_1 = \dfrac{1}{\tau + \tau_t}$.

## 3. Coupled LB-DEM model

### 3.1 General

The Lattice Boltzmann approach has the advantage of accommodating large particle sizes and the interaction between the fluid and the moving particles can be modelled through relatively simple fluid - particle interface treatments. Further, employing the Discrete Element Method (DEM) to account for the particle/particle interaction naturally leads to a combined LB - DEM solution procedure. The Eulerian nature of the Lattice Boltzmann formulation, together with the common explicit time step scheme of both the Lattice Boltzmann and the Discrete Element, makes this coupling strategy an efficient numerical procedure for the simulation of particle-fluid systems[13]. The LB-DEM coupling system is a powerful fundamental research tool for investigating hydro-mechanical physics in porous media flow [14]. To capture the actual physical behaviour of a fluid-particle system, the boundary condition between the fluid and the particle is modelled as a non-slip boundary condition, i.e. the fluid near the particle should have a similar velocity as the particle boundary. The solid particles inside the fluid are represented as solid lattice nodes. The discrete nature of lattice will result in stepwise representation of the surfaces [15]. A very small lattice spacing is adopted to obtain smoother boundaries.

The smallest DEM grain in the system controls the size of the lattice. In the present study, a very fine resolution of $d_{min}/h = 10$ is adopted. That is, the smallest grain with a diameter $d_{min}$ in the system is discretized into 100 lattice nodes ($10h$ x $10h$). This provides a very accurate representation of the interaction between the solid and the fluid nodes.

When combining the Discrete Element modelling of grain interactions with the lattice Boltzmann formulation, an issue arises. That is, there are now two time steps: $\Delta t$ for the fluid flow and $\Delta t_D$ for the particle movements. Since $\Delta t_D$ is normally smaller than $\Delta t$, $\Delta t_D$ is slightly reduced to a new value $\Delta t_s$ so that $\Delta t$ and $\Delta t_s$ have an integer ratio $n_s$:

$$\Delta t_s = \Delta t / n_s \quad \& \quad (n_s = [\Delta t/\Delta t_D]+1) \tag{11}$$

This results in a subcycling time integration for the Discrete Element part. At every step of the fluid computation, $n_s$ sub-steps of integration are performed for DEM using the time step $\Delta t_s$. The hydrodynamic force is unchanged during this sub-cycling.

## 3.2 Modelling Permeability

In DEM, the grain – grain interaction is described based on the contact interactions. In a 3D granular assembly, the pore spaces between grains are interconnected, whereas in 2-D assembly, the grains are in contact with each other that result in a non-interconnected pore-fluid space. This results in a no flow condition in a 2-D case (see Figure 2). To overcome this difficulty, a reduction in radius is assumed only during LBM computations (fluid and fluid – solid interaction), which is called the hydrodynamic radius. The hydrodynamic radius allows interconnected pore space through which the surrounding fluid can flow (hydrodynamic radius $r = 0.7\,R$ to $0.95\,R$, where '$R$' is the grain radius). The hydrodynamic radius is used only during the LBM computations, and has no effect on the grain – grain interactions computed using DEM. Different values of macroscopic permeability can be obtained for any given initial packing by varying the hydrodynamic radius of the grains, without having to change the actual granular packing. This introduces a new parameter into the system. In a physical sense, a hydrodynamic radius represents the three-dimensional permeability of a granular assembly simulated as a two-dimensional geometry.

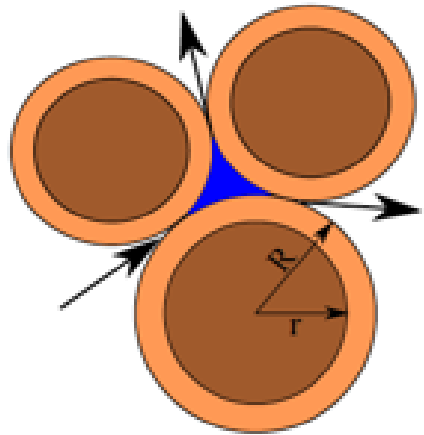

Figure 2: Schematic representation of the hydrodynamic radius in LBM-DEM computation.

In order to understand the relation between the hydrodynamic radius and the permeability of a granular assembly, permeability tests are performed by varying the hydrodynamic radius $r$ as 0.7 $R$, 0.75 $R$, 0.8 $R$, 0.85 $R$, 0.9 $R$ and 0.95 $R$. The permeability values obtained are normalized by the square of the average grain diameter following the Carman-Kozeny equations [16]. The comparison of normalised permeability from the 2D LB-DEM simulations with the Carman-Kozeny equations for spherical and cylindrical grain assembly for different porosities are presented in Figure 3. It can be observed from the figure that the permeability decreases drastically as the radius is decreased from 0.7R to 0.95R. The granular assembly is almost impermeable for a hydrodynamic radius of 0.95R. The normalized permeability is found to match the qualitative trend of the Carman-Kozeny equations.

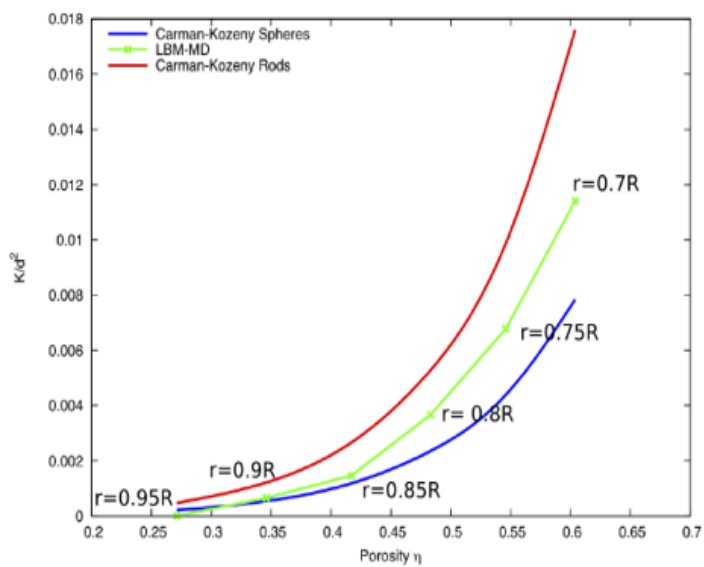

Figure 3: Relation between permeability and porosity for different hydrodynamic radius and comparison with the analytical solution.

# 4. Granular column collapse on slopes in fluid

4.1 Problem Definition

In this study, a 2D polydisperse system ($d_{max}/d_{min} = 1.8$) of circular discs in fluid was used to understand the behaviour of granular flows on inclined planes. As shown in Figure 4, an inclined gravity direction was varied to examine the slope angle effect. The soil column was modelled using 1000 discs of density 2650 kg m$^{-3}$ and a contact friction angle of 26°. The aspect ratio '$a$' is defined as the ratio of the initial height ($H_i$) to the width ($L_i$) of the column. A granular column of aspect ratio '$a$' of 0.8 was used. The collapse of the column was simulated inside a fluid with a density of 1000 kg m$^{-3}$ and a kinematic viscosity of $1 \times 10^{-6}$ m$^2$s$^{-1}$. The choice of a 2D geometry has the advantage of cheaper computational effort than a 3D case, making it feasible to simulate very large systems. To model 3D permeability nature of spheres as 2D discs, a reduction in radius: a hydrodynamic radius $r = 0.9R$ was adopted only for LBM computations, as described earlier. Dry column collapse was also performed to study the effect of hydrodynamic forces on the runout distance. The runout distance is normalised with respect to the initial width of the column. The granular column collapse was allowed to flow down slopes of varying inclinations (0°, 2.5°, 5° and 7.5°).

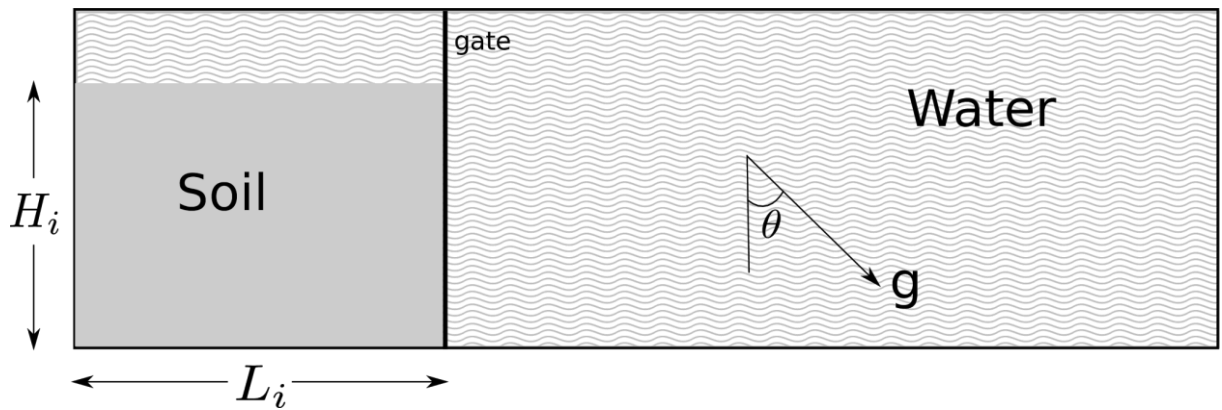

Figure 4: Underwater granular collapse set-up

4.2 Collapse of a dense granular column

Figure 5 shows the snapshots of the flow evolution of a dense granular column (aspect ratio 0.8), which has an initial packing density of $\Phi = 83\%$, on a 5° slope. The failure begins at the toe end of the column, and the shear-failure surface propagates into the column at an angle of about 45 - 50° The failure is due to collapse of the flank. Force chains can be observed in the static region of the column. Once the material is destabilised, the surface of the flowing granular mass interacts with the surrounding fluid, resulting in the formation of turbulent vortices. The vortices are formed only during the horizontal acceleration phase and interact with the grains at the surface resulting in an irregular free surface. As the granular material ceases to flow, force chains restart to develop at the flow front, revealing consolidation of the granular mass with an increase in the shear resistance. The vortices formed during the collapse start to raise above the settled granular mass.

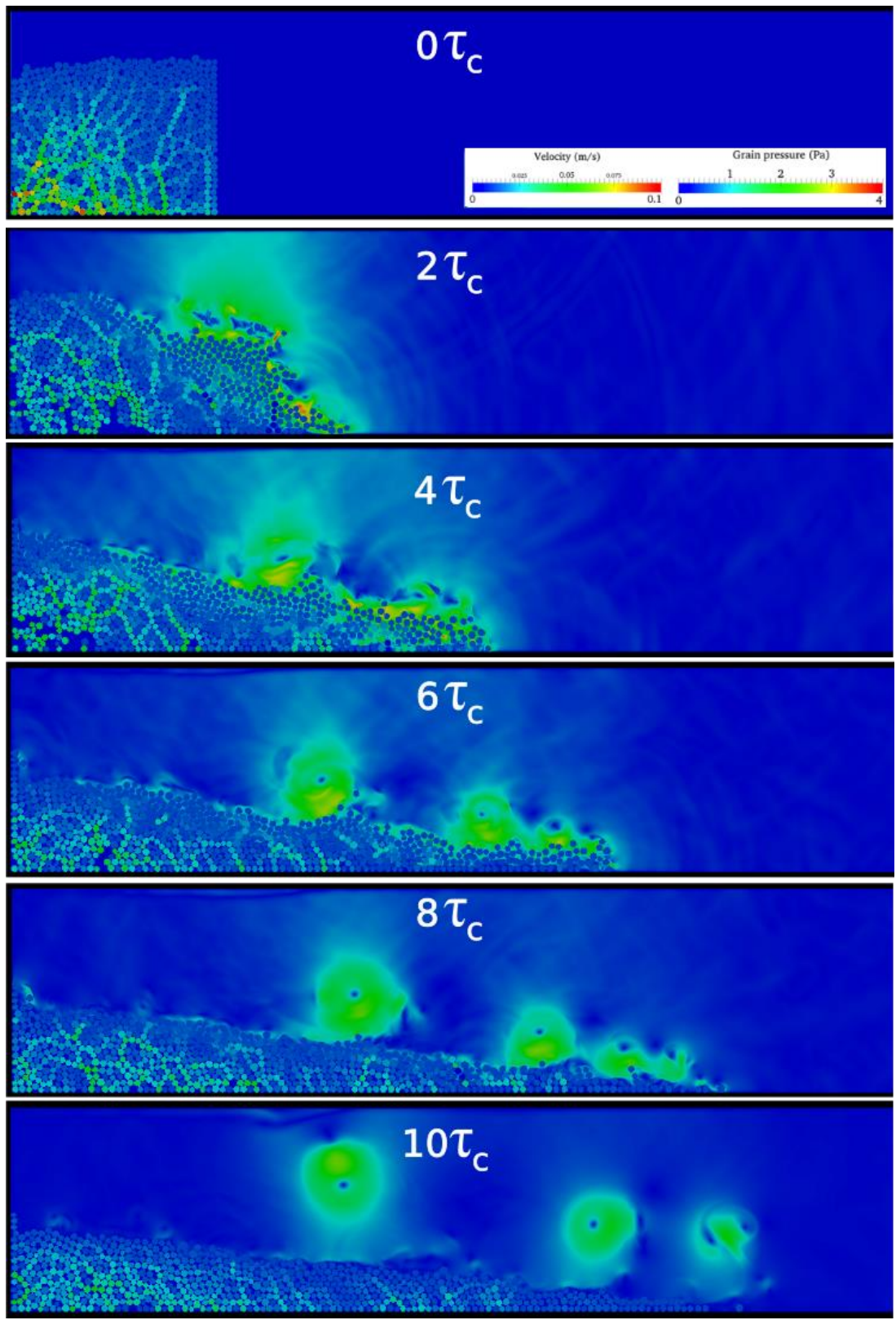

Figure 5: Evolution of a dense sand column (aspect ratio 0.8) on a slope of 5°

The runout distance is measured by tracking the farthest particle that is still in contact with the main granular mass. The normalized runout distance is measured as $\Delta L = (L_f - L_i)/L_i$, where $L_f$ is the runout distance at a given time and $L_i$ is the initial width of the column. Figure 6a shows the evolution of the normalised runout with time for the collapse of a dense sand column in the dry and submerged cases for varying slope angles. For the horizontal axis, the time is normalised as $t / \tau_c$, where $\tau_c$ is defined as the critical time of the dry granular column collapse to be fully mobilised on a horizontal plane ($\tau_c = \sqrt{h/g}$)[17]. For all slope angles (including the 0° case), the runout distances in the dry case are longer than those observed in the submerged cases. The difference in the runout between the dry case and the submerged case decreases with increase in slope angle. At a slope angle of 5°, the difference is the smallest. The reason for this is explained by examining the kinetic energy of the moving granular body, as discussed next.

Figure 6b shows the evolution of the normalised kinetic energy with time for the dry and submerged collapses with different slope angles. In general, the submerged granular collapse cases show about half the peak kinetic energy of the dry collapse cases, indicating energy dissipation by the drag force at the interface between the granular body and the free water. The critical time, which is the time to reach the peak kinetic energy, for the submerged granular collapse cases is about 3 times slower than the dry collapse cases on a horizontal plane. Within the submerged cases, the longest sustained peak kinetic energy is observed for the slope angle of 5°. This explains the smaller difference in the runout distance between the dry and submerged cases for the slope of 5°. A sustained peak kinetic energy is typically observed during hydroplaning of a thin flowing layer, as discussed in the later section on hydroplaning.

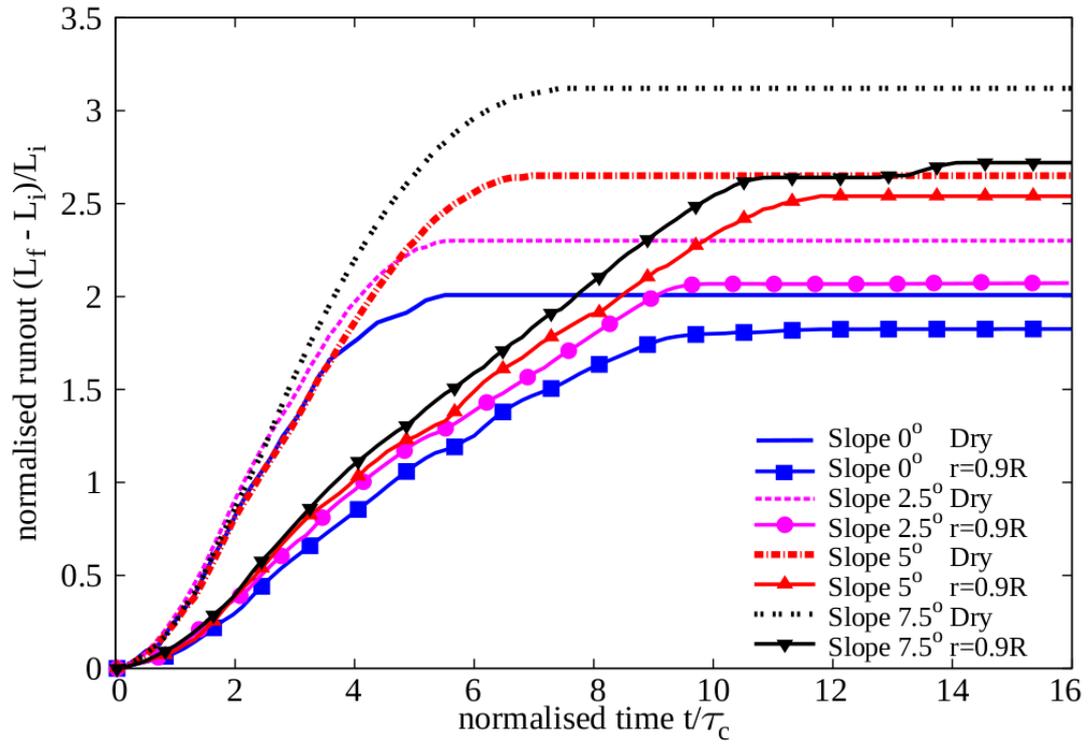

(a) Evolution of run-out with time.

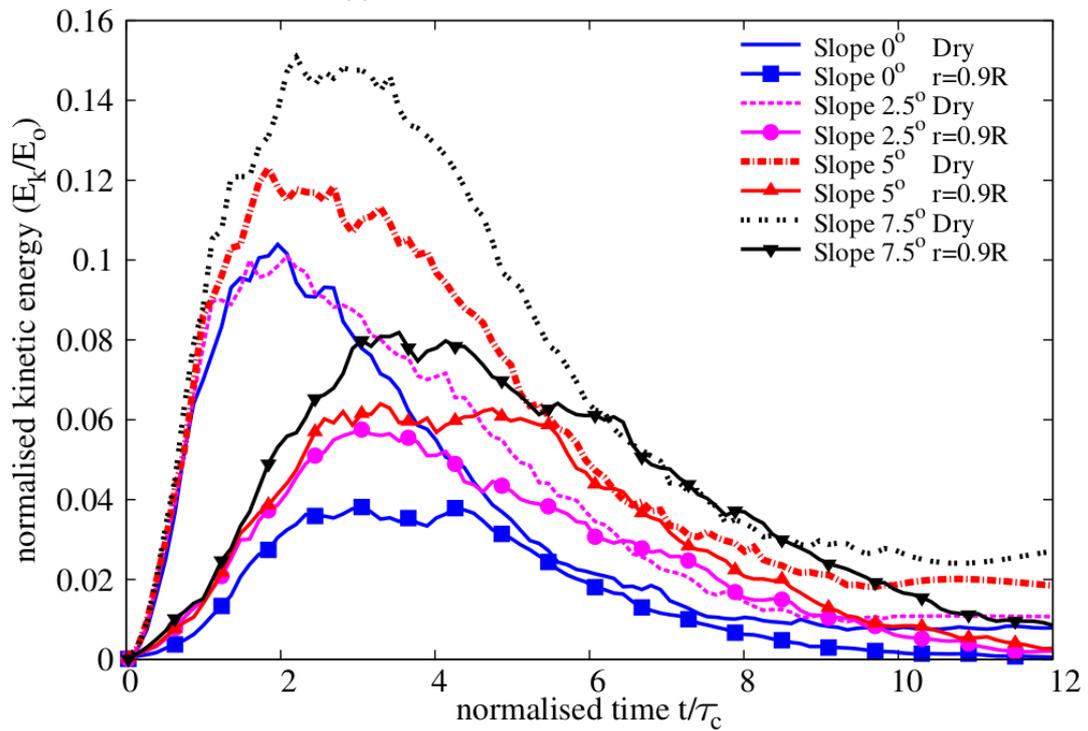

(b) Evolution of kinetic energy with time.

Figure 6 Evolution of runout and kinetic energy with time (dense case).

The snapshots of the dense granular column collapse down slopes of varying inclinations at the time of $t = 3\tau_c = 3\sqrt{h/g}$) are shown in Figure 7. This corresponds to the time at which peak kinetic energy is achieved, i.e., the flow is fully mobilised. The colour contours in the fluid phase are the flow velocities, whereas those in the solid phase are the interparticle forces. Fast velocity or large forces is shown in red, whereas nearly zero velocity or force is shown in blue. With increase in slope angle, the amount of material destabilised increases, which increases the surface area of the granular mass interacting with the surrounding fluid. In turn this increases the hydrodynamic drag on the granular surface.

Figure 8 shows the evolution of flow front for different slope angles by zooming into the front part of the granular mass shown in Figure 7. The fluid velocity increases (shown in red corresponding to 0.1 m/s) with increase in the slope angle along the surface of the granular mass. Increase in fluid velocity indicates momentum transfer between the flowing granular mass and the surrounding fluid, i.e., drag force. The 7.5° slope angle case experiences high hydrodynamic drag, which results in a shorter runout distance than the dry case.

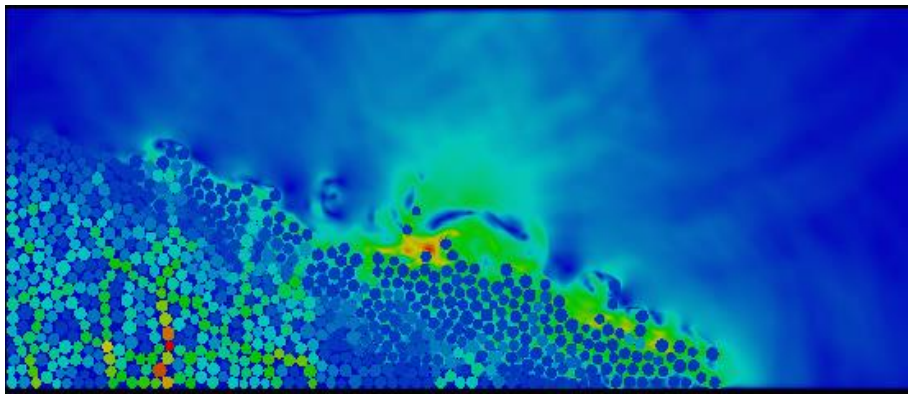

(a)     Slope angle 2.5°

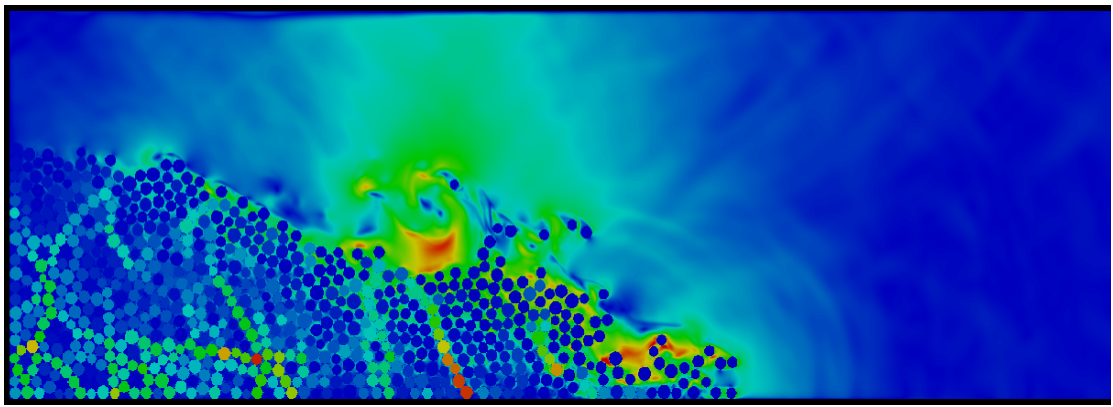

(b)     Slope angle 5°

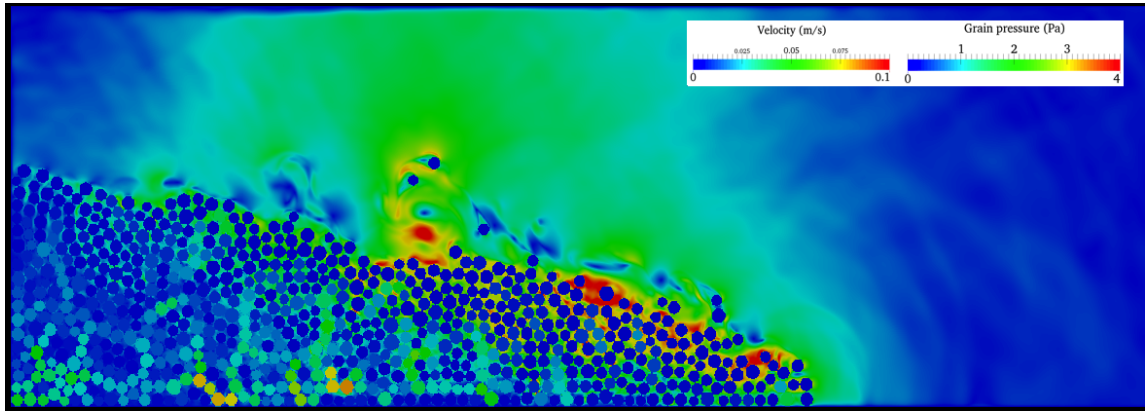

(c)     Slope angle 7.5°

Figure 7 Flow morphology at t = 3$\tau_c$ for different slope angles (dense).

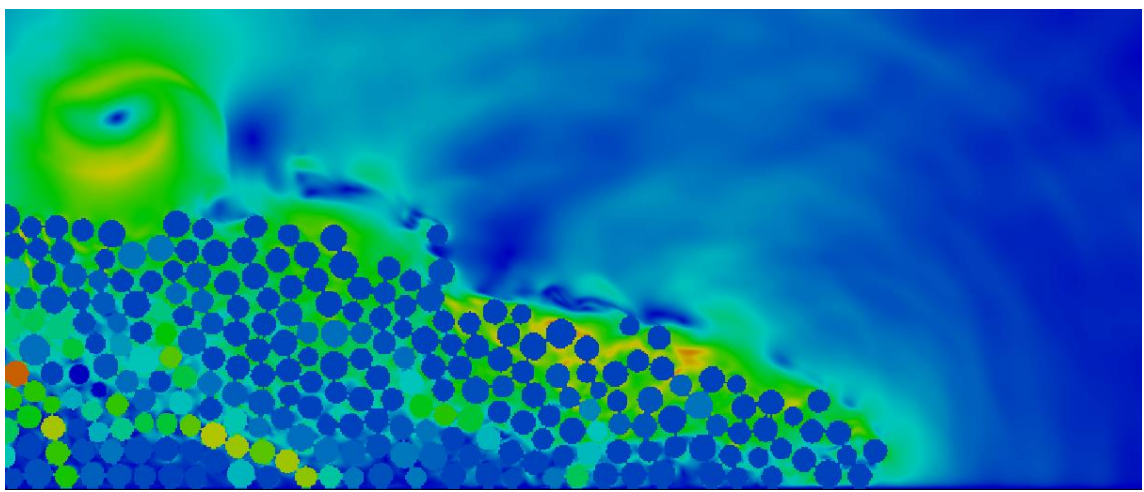

(a)     Slope angle 2.5°

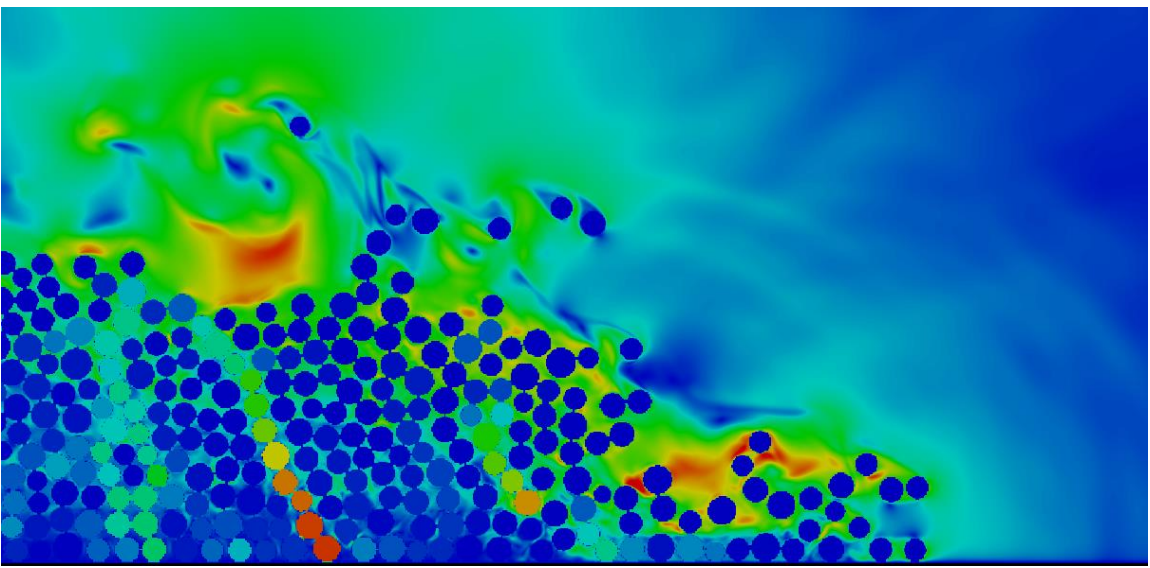

(b)     Slope angle 5°

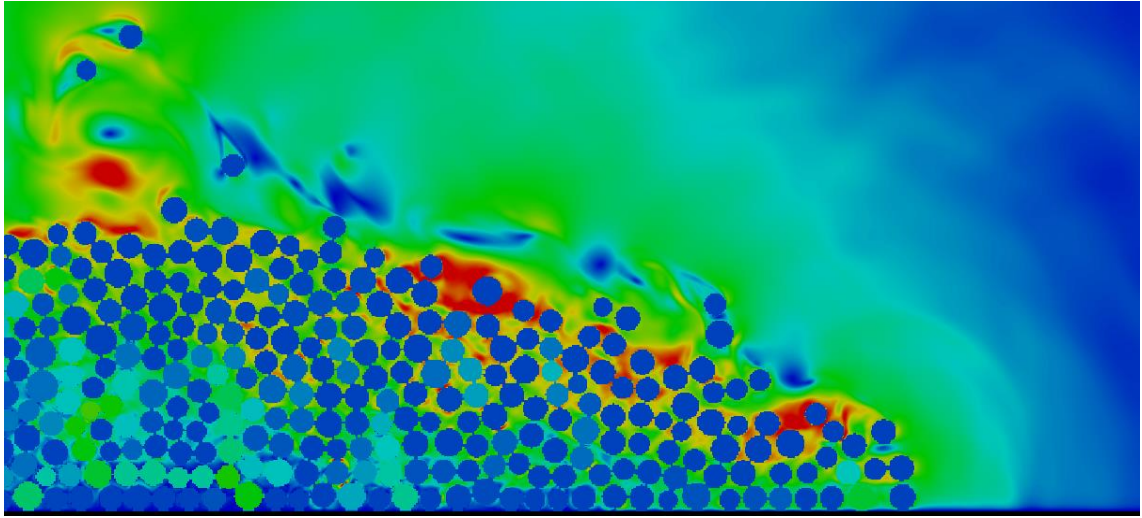

(c)    Slope angle 7.5°

Figure 8 Evolution of flow front at t = 3τ$_c$ for different slope angles (dense).

From the interparticle force distributions presented in Figure 7, a central static core region can be identified for all slope angles. However, with increase in slope angle, the failure plane angle from the toe end of the column decreases, which increases the amount of mobilised mass. This can be observed by the decrease in the volume of the static region in Figure 7. With increase in slope angle, the length of the failure plane increases. Figure 9 show the pore pressure along the failure planes for different slope angles at the initiation stage, where a normalised length of 0 refers to the wall side of the column and a value of 1 denotes the toe of the column at the initial step. As the material starts to flow, negative excess pore pressure develops inside the granular mass, which in turn increases the particle interaction forces (or effective stress). The 7.5° slope case has the largest negative excess pore pressure along the failure plane due to large shear mobilization of a larger granular body. As the material is dense and has dilative tendency in shearing, more negative excess pore pressures develop. The 2.5° and 5° slope cases also show development of negative excess pore pressure but the development is more localized near the toe of the column (normalised length ~0.8) indicating the initiation of the failure plane at the toe of the column. The 7.5° slope case shows positive pore pressures at the toe of the column, indicating flow progression, i.e., the granular mass has overcome the negative pore pressure and has started to flow. As the column starts to flow, positive pore pressure can be observed in front of the toe of the column (normalised length 1), this can be observed at all slope angles.

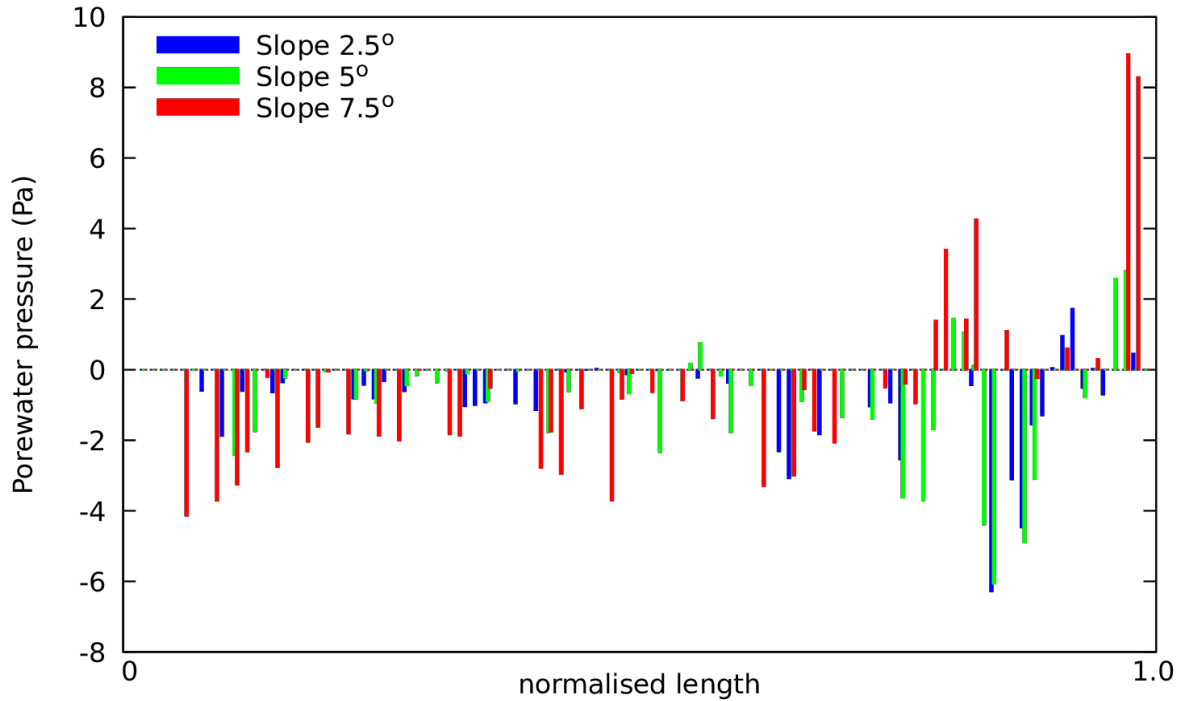

Figure 9: Pore pressure along the failure plane at collapse initiation for different slope angles. Normalised length is defined as the length of the failure plane.

## 5. Effect of initial density - Loose versus Dense granular column

Topin et al. [4] observed from CFD-DEM simulations of granular collapse on a horizontal plane in different viscous fluids that, for a given initial geometry, the run-out distance in the dry case is significantly higher than the submerged case, an observation similar to the experimental results of Cassar et al. [18]. For a given geometry, the initial volume fraction of the granular mass has a significant effect on the morphology of the granular deposits in fluid [19], [20]. In the dry case, inertia is responsible for the enhanced mobility at steeper slopes. In the submerged cases, however, the grain-inertial effects remain negligible because of the predominance of viscous effects on the granular dynamics [4]. This could explain the importance of the initial volume fraction on the runout behaviour of submerged granular columns. In this study, we examine the effect of initial packing density on the runout behaviour at various slope angles in both dry and submerged cases.

In order to understand the influence of the initial packing density on the runout behaviour, a collapse of loose sand column (initial packing density $\Phi = 79\%$) was simulated at different slope angles and the results were compared to the dense sand column cases (initial packing density $\Phi = 83\%$), which were presented earlier.

Figure 10a shows the evolution of normalised runout distance with time for the initially loose granular column cases. In contrast to the dense granular column cases, the loose granular columns show longer runout distance in the submerged cases.

Figure 10b shows the evolution of kinetic energy with time for different slope angles for the loose granular column collapse cases. The difference in the peak kinetic energy between the dry and submerged cases is about 20 - 40% compared to difference in the peak kinetic energy of almost half in the dense case. The difference in the peak kinetic energy between the dry and the submerged case is the lowest (about 20%) for a slope angle of 5°. The sustained peak kinetic energies can be observed in the submerged case in contrast to their dry counterparts, which in turn results in longer runout distance in the submerged case compared to the dry case.

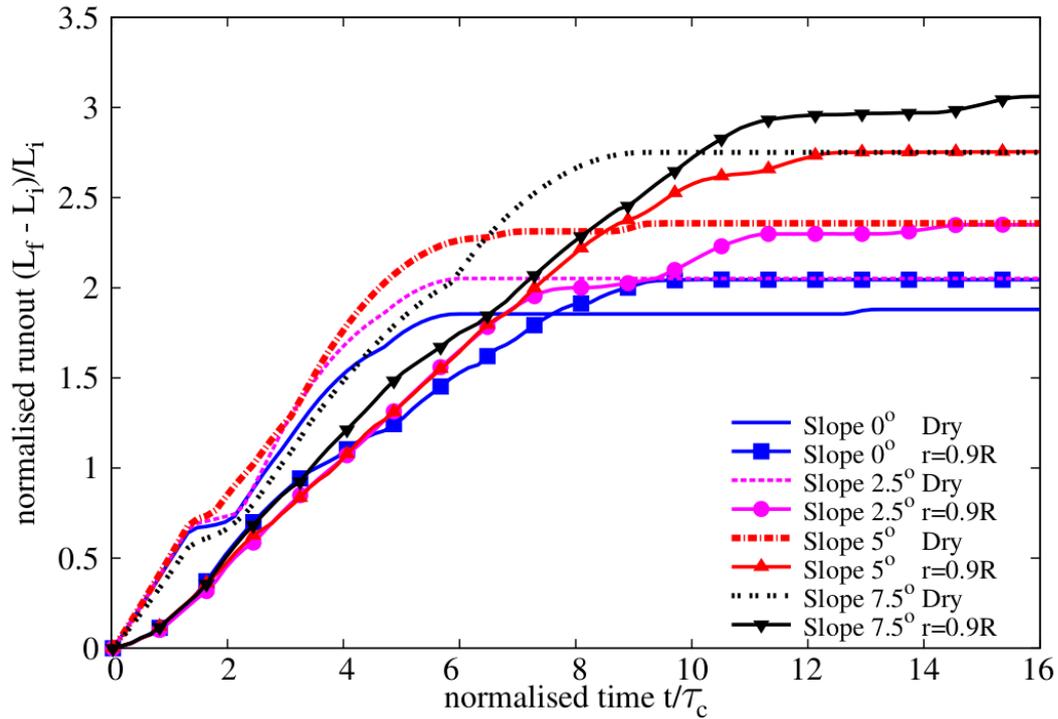

(a) Evolution of run-out with time.

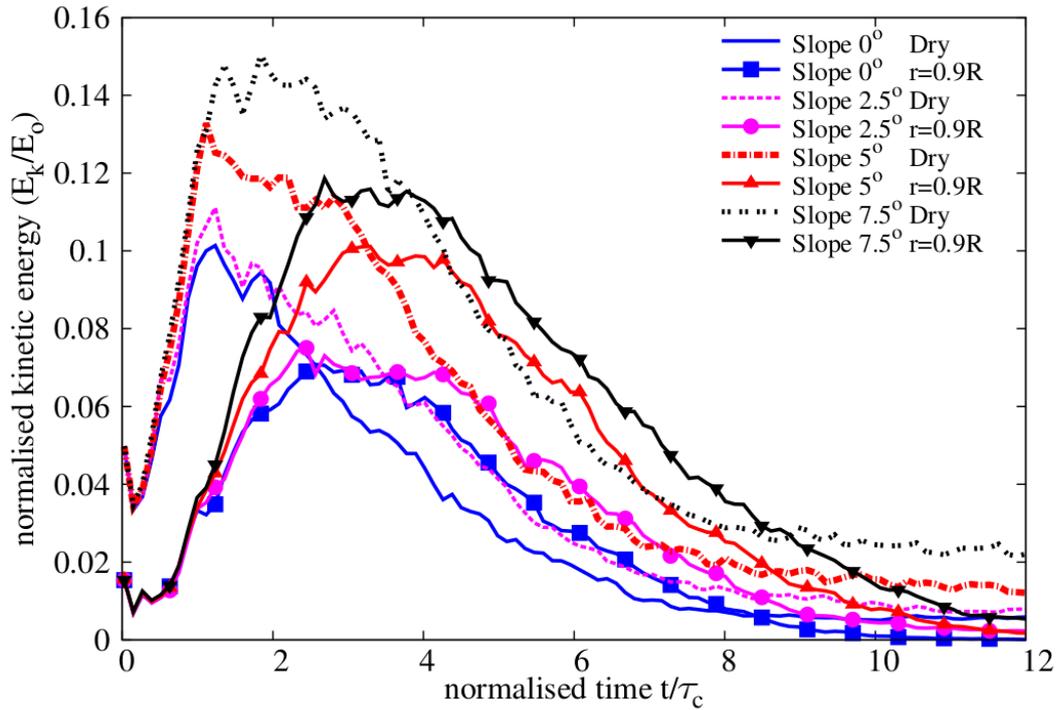

(b) Evolution of kinetic energy with time.

Figure 10 Evolution of runout and kinetic energy with time for different slope angles (loose case).

Figure 11a shows the final runout distance for the dense granular column cases, while Figure 11b shows that for the loose granular column cases. As the slope angle increases, the final runout distance increases in both dry and submerged cases. For the dense cases (Fig. 8a), the runout in the dry case is larger than the submerged cases for all slope angles. As discussed earlier, as the slope angle increases, the drag force experienced by the dense granular column plays a dominant role on the runout behaviour. This results in an increase in the difference between the dry and the submerged cases with increase in the slope angle. For the loose granular columns (Fig. 8b), on the other hand, the columns in the submerged cases flow longer than those in the dry cases, which is the opposite trend observed in the dense cases. Figure 12 shows that for all slope angles, the runout distance in the submerged loose granular collapse cases is greater than the dense collapse cases. The difference in the magnitude of the runout distance increases with slope angle.

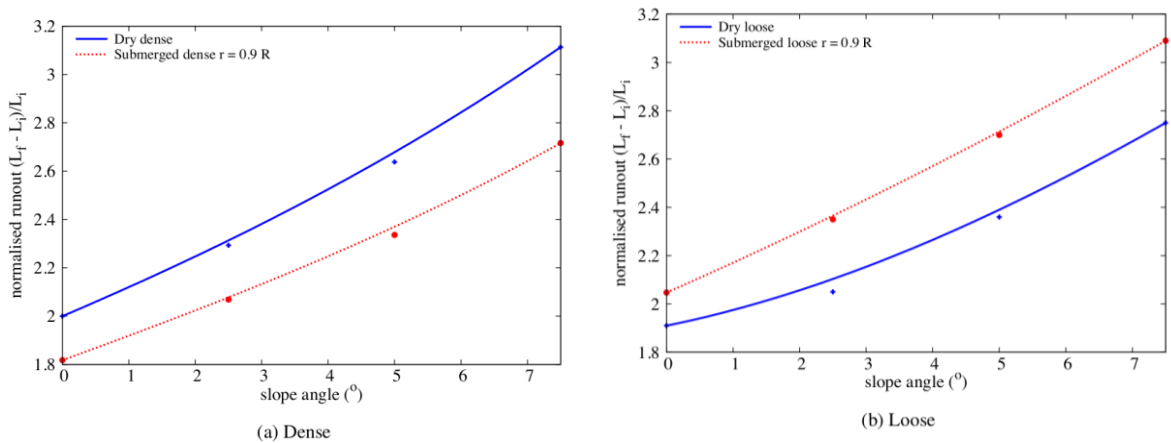

Figure 11 Comparison between dry and submerged granular column on the effect of slope angle on the runout distance (Dense and Loose).

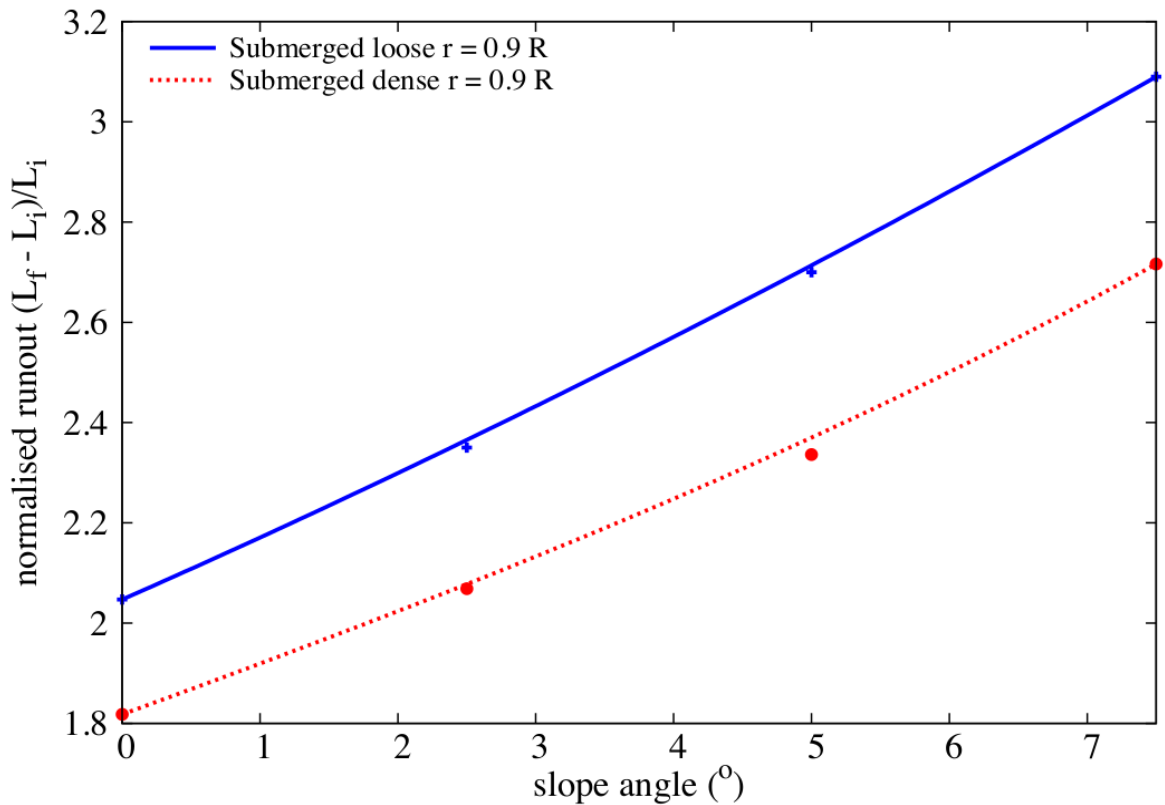

Figure 12 Effect of slope angle on the runout behaviour for different initial packing density.

## 6. Mechanisms of granular column collapse in fluid

The difference in the runout behaviour between the dense and loose granular columns at various slope angles suggests difference in runout mechanism. The collapse of a granular column involves three stages: an initiation stage characterised by a distinct failure surface, a runout phase characterised by horizontal acceleration which involves formation of eddies, and the final settlement phase. The mechanisms observed in each phase are discussed next.

### 6.1 Initiation phase

Figure 13 compares the initial evolution of runout between the loose and the dense cases (t = 0 to $3\tau_c$). For all slope angles, the loose granular columns evolve faster than their dense counterpart. When a granular material is sheared in the submerged conditions, it generates negative pore water pressure initially due to the undrained conditions. In the undrained conditions, the dilation movement (volume expansion) of the granular assembly by shearing self-generates negative excess pore pressure inside the pores. The fluid inside the pores does not have time to seep in or out from the free water outside the column by the internally generated pressure gradient between the pore space and the outside free water. The negative excess pore pressures produce larger interparticle forces by bringing the grains together. Macroscopically this results in increase in the mean effective stress and provides large shear resistance temporarily until the negative excess pore pressure dissipates with time. For example, Figure 9 showed large negative pore pressures for the dense cases. The length of the failure plane increases with the slope angle, thereby increasing the volume of body with negative pore pressure that needs to be dissipated before the column starts to move. Overcoming the region of large negative pore pressure translates to smaller kinetic energy and eventually a shorter runout distance for the dense granular columns.

The degree of dilation upon shearing is smaller in the loose granular column case than the dense case. As shown in Figure 14, positive pore pressures can be observed along the failure plane for all slope angle cases. Positive pore pressures imply that the grains are pushing apart by increasing pore pressure during the initial phase of the collapse. The intergranular forces (or effective stress) are reduced and consequently the shear resistance decreases. This results in faster initiation and runout behaviour for the loose granular columns compared to the dense granular columns.

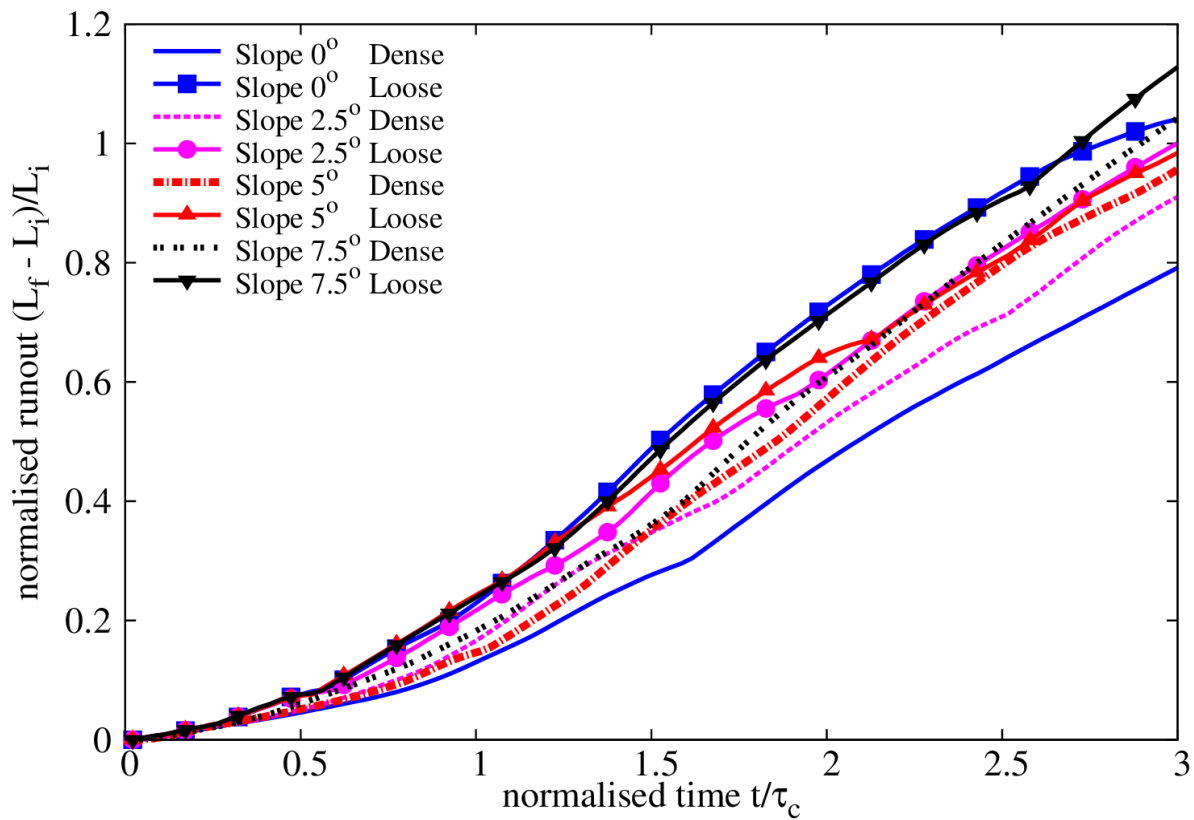

Figure 13. Initial runout evolution comparison between the loose and dense submerged cases for all slope angles.

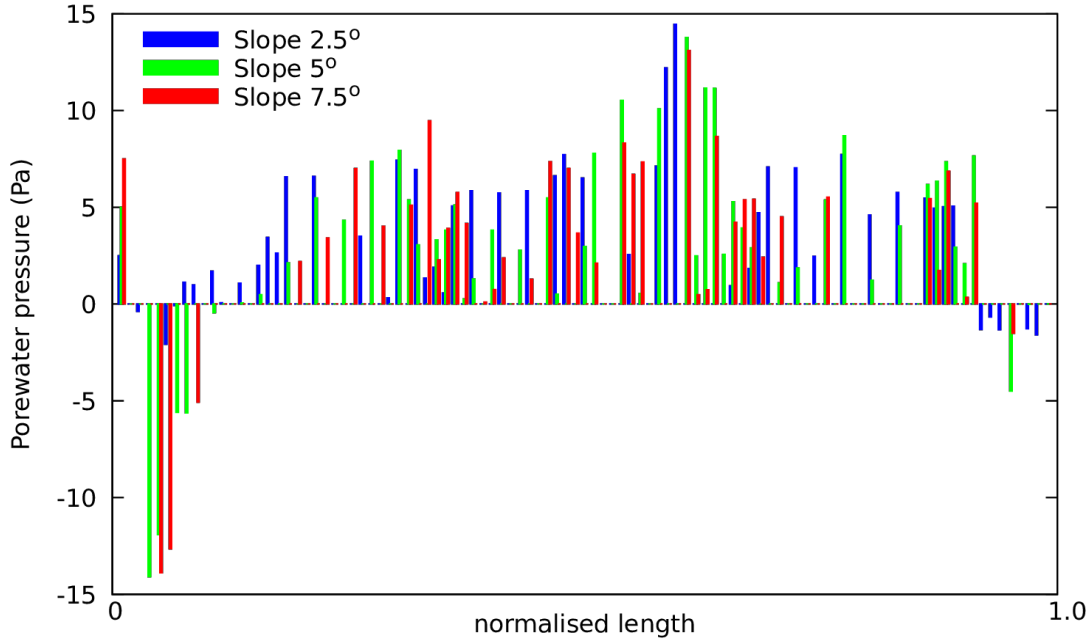

Figure 14: Pore pressure along the failure plane at collapse initiation for different slope angles. Normalised length is defined as the length of the failure plane.

## 6.2 Runout phase

In the second phase, the initial vertical collapse motion is converted to horizontal acceleration. The snapshots of the loose granular collapse cases at t = $3\tau_c$ for different slope angles are presented in Figure 18. The flow is fully mobilised at $3\tau_c$, when the kinetic energy is at its peak. In the submerged cases, the dynamics of the granular flow are controlled by the following three factors: (a) hydroplaning, defined as the loss of friction between the flowing material and the bottom surface due to the presence of water, (b) water entrainment at the front of the flowing mass, which results in a decrease in the density of the flowing granular mass, and (c) the interaction of the surface of granular flow with the surrounding fluid resulting in formation of eddies and drag effects. The role of water on the dynamics of the dense and loose granular columns is discussed in this section.

### 6.2.1 Hydroplaning

Hydroplaning refers to the loss of friction, which occurs due to the entrapment of water between the granular mass and the slope, which results in zero effective stresses. Figure 15 shows the effective stress of the flow front for a distance of ~15d for the slope angle 5° case. The effective stress is computed from the contact forces acting between the grains and the bottom surface. The horizontal axis is normalised to the maximum length of ~15d from the flow front and the vertical axis is normalised to the initial maximum effective stress. The average effective stress at the bottom of the flowing granular mass over 5 time-steps is presented to avoid fluctuations in the stresses. The loose granular columns have zero average effective stress at the flow front (~7.5d from the front). In contrast, the dense granular columns that have positive effective stress at the sliding interface, which creates frictional forces that work against the flow movement. The loose granular flow tends to entrain more water at the base of the flow front, creating a hydroplaning surface. This results in a longer runout distance. This corroborates the influence of hydroplaning on the loose granular columns which show higher sustained peak kinetic energy in comparison with the dense granular columns.

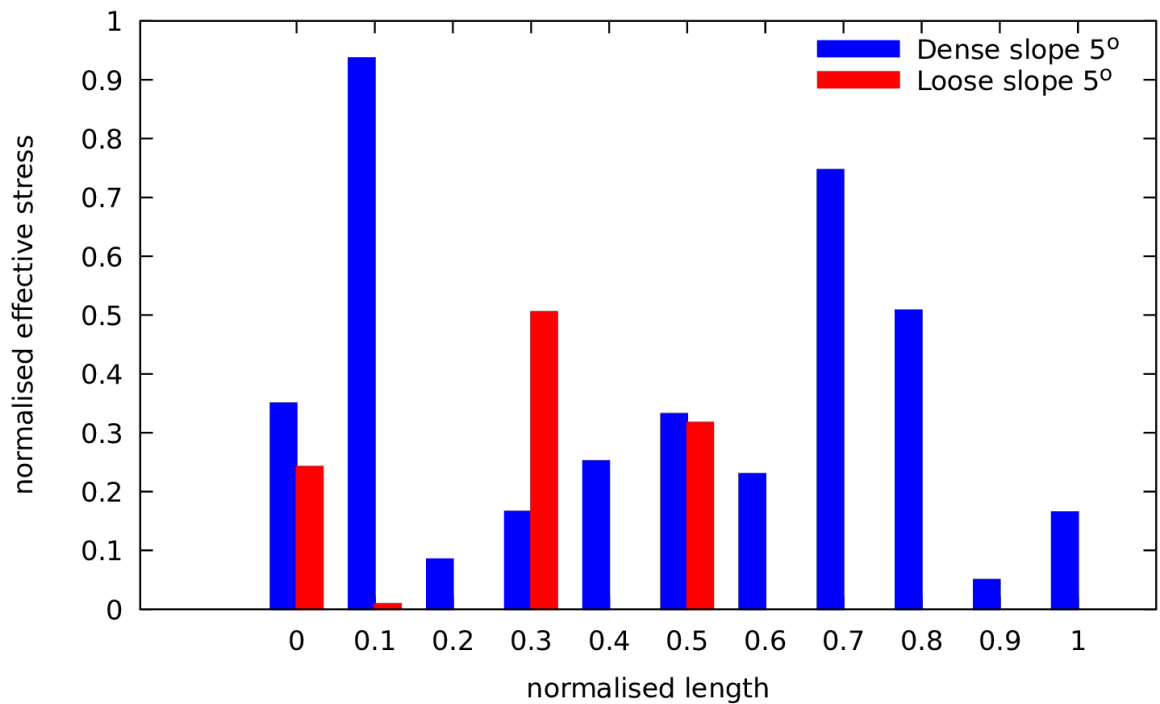

Figure 15: Effective stress at the flow front (~15d) between the dense and loose submerged cases for a slope angle of $5°$ at time t = $3\tau_c$

### 6.2.2 Water entrainment

Comparing the snapshots of the granular flow at time t = $3\tau_c$ for the dense column cases (Figure 8) and the loose (Figures 6) column cases, it can be seen that the overall shapes of the flow front are different. Figure 16 shows the outline of the flowing mass for various slope angles for both cases. Figure 17 shows the boundary of the flowing mass and its evolution with time for the slope angle 5° case. Both figures show that the dense granular columns exhibit a more acute angle of attack (flow front) in contrast to a more parabolic profile for the loose granular columns. This difference in angle of flow front changes the ratio of amount of water entering the flowing granular mass to the water interacting with the surface of the granular mass. Figure 19 shows the horizontal velocity of the fluid and streamlines at the flow front of the dense and loose granular columns for the slope angle of 5°. The density of streamlines entering the granular front is significantly higher for the loose column case, in comparison to the dense column case. This clearly shows the effect of the shape of flow front on water entrainment. More water entrainment results in increase in pore space (or decrease in solid fraction) at the flow front, which in turn reduces the interparticle shear resistance.

The water entrapped at the flow front results in a drop in the density of the flowing mass, which causes smaller contact forces, i.e. lower effective stress enabling lubrication between particles. Figure 20a shows the change in the packing density of the overall granular mass. In the loose column cases, the amount of water entering the front decreases the packing density, making the mass looser to flow (Figure 20a at t = $3\tau_c$). More water is entrained with increasing slope angle, resulting large decrease in packing density. In the dense column cases, the packing density also decreases with time mainly due to shear induced dilation of the overall moving mass as the negative excess pore pressure developed at the beginning dissipates with time.

Figure 20b shows the evolution of Froude's number for the submerged cases. For the same thickness and the velocity of the flow, the loose granular flow has a smaller density and hence a higher Froude's number than the dense granular flow, resulting in a higher probability of hydroplaning and more water entrainment. Fast moving granular flows undergo a motion-induced self-fluidisation process due to flow front instabilities setting on at large values of Froude's number, which are responsible for extensive entrainment, and longer time between collisions of soil grains. Self-fluidisation results in enhanced mobility of the solids, causing an inviscid flow. Bareschino et al. [21] observed similar instabilities in experiments on sub-aerial rapid granular flows. Harbitz [22] experimentally observed that hydroplaning occurs above a critical value of densimetric Froude number of 0.4. At time t = $3\tau_c$, the dense column has a densimetric Froude number just above 0.4, while for the loose column the value is 0.65. Our grain-scale numerical simulations also show a similar finding to the experimental observations of hydroplaning at densimetric Froude number above 0.4.

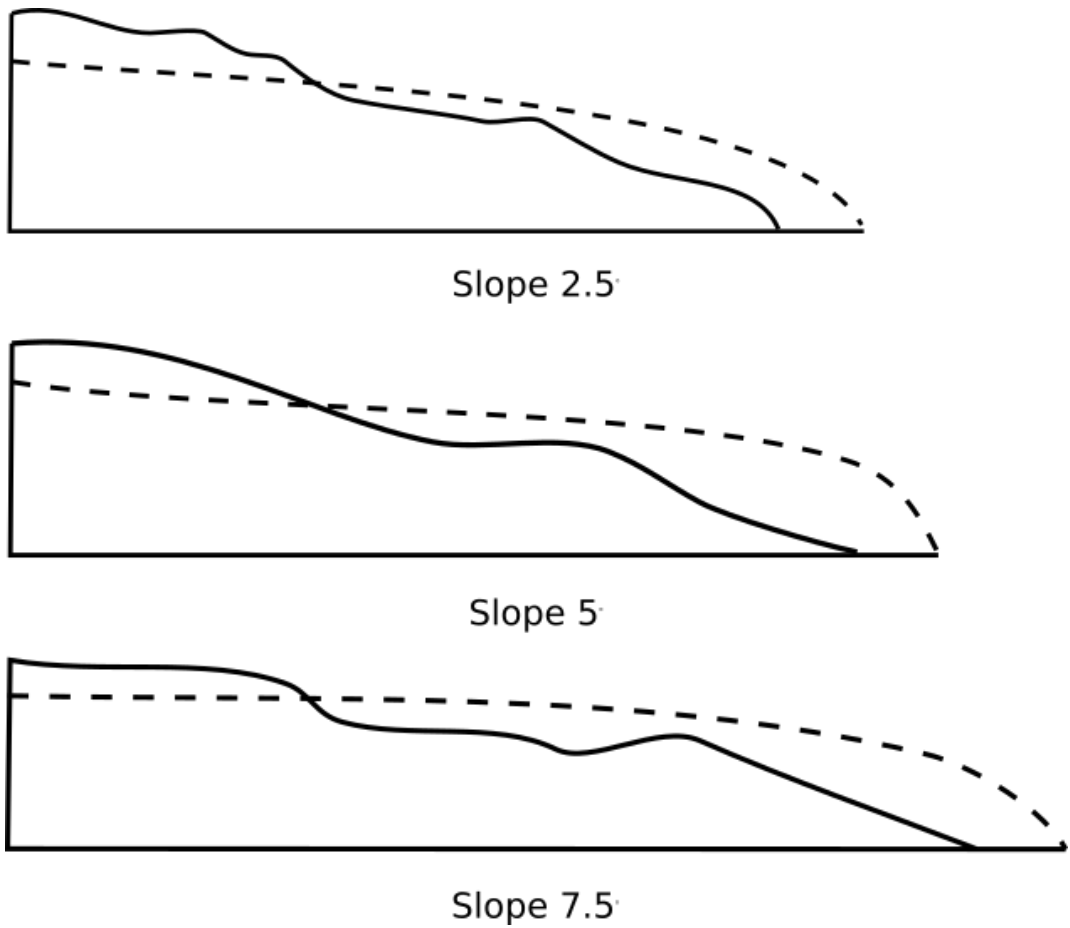

Figure 16: Slope profile outline for dense (solid line) and loose (dashed line) at time $t = 3\tau_c$

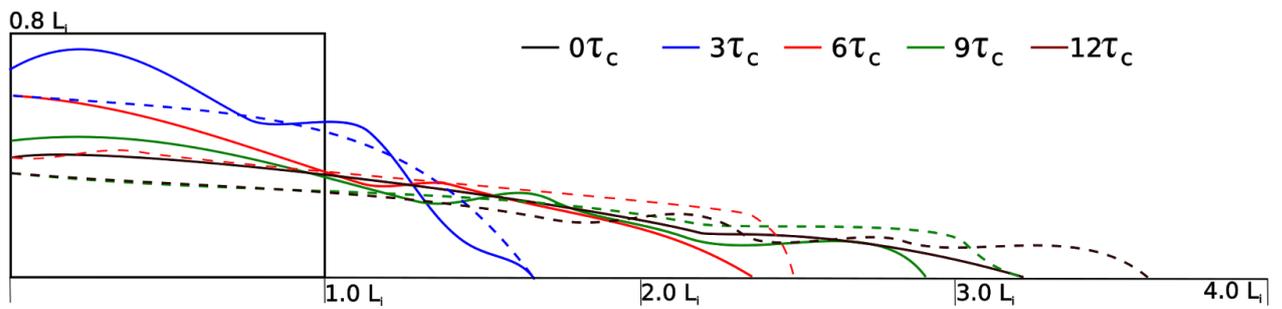

Figure 17: Profile outline evolution for dense (solid line) and loose (dashed line) cases for a slope of 5°

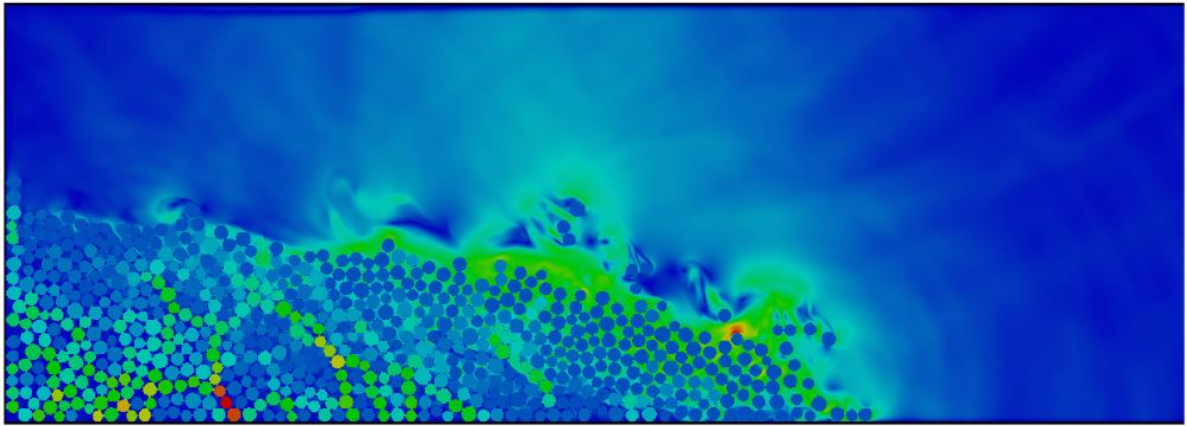

(a) Slope 2.5 °

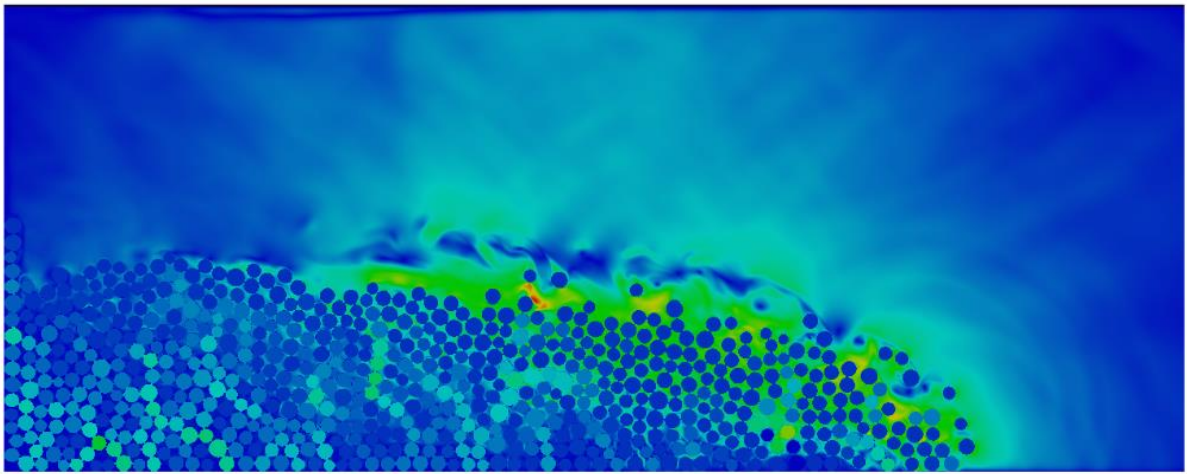

(b) Slope 5.0 °

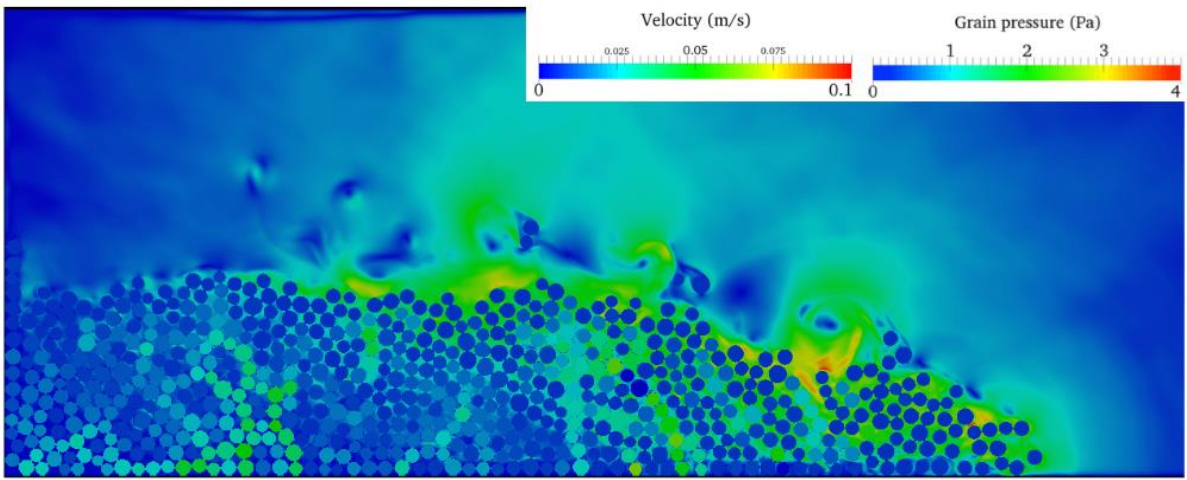

(c) Slope 7.5 °

Figure 18: Flow morphology at time t = 3τ$_c$ for different slope angles (loose granular columns).

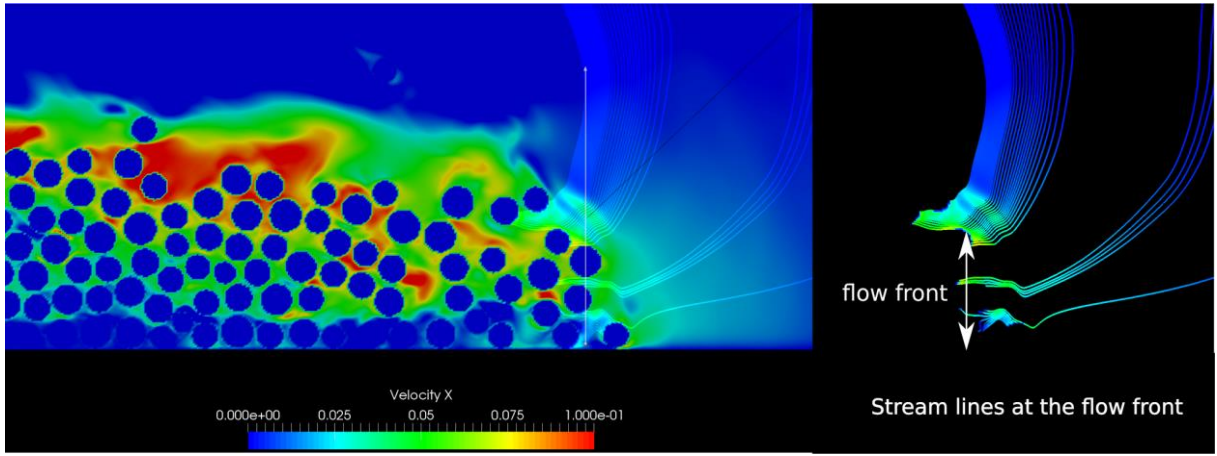

(a) streamlines at the flow front dense case (contours show velocity of fluid entrainment at the flow front)

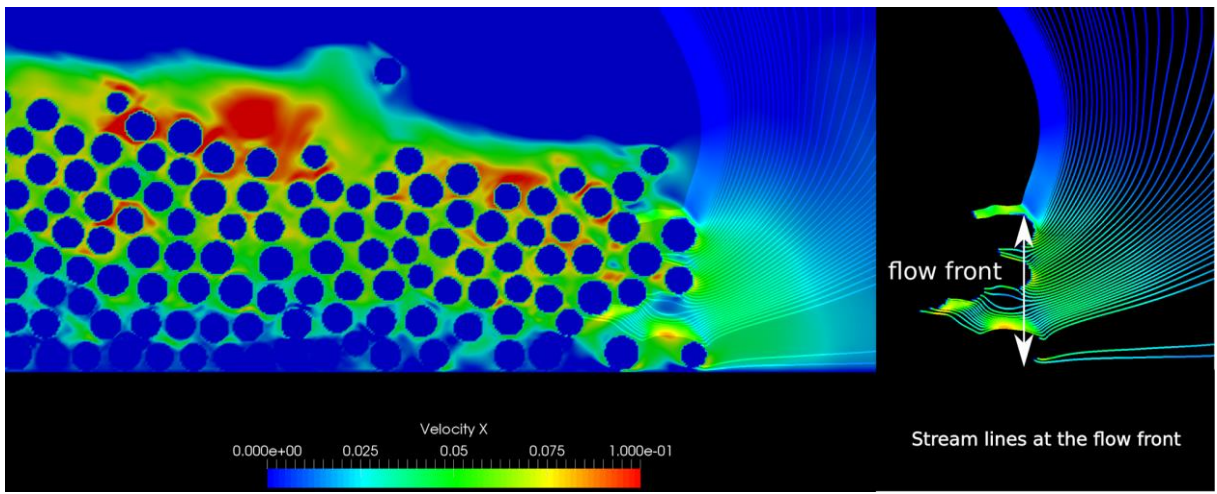

(b) streamlines at the flow front for the loose case (contours show velocity of fluid entrainment at the flow front)

Figure 19: Water entrainment for a slope of 5° (dense and loose cases).

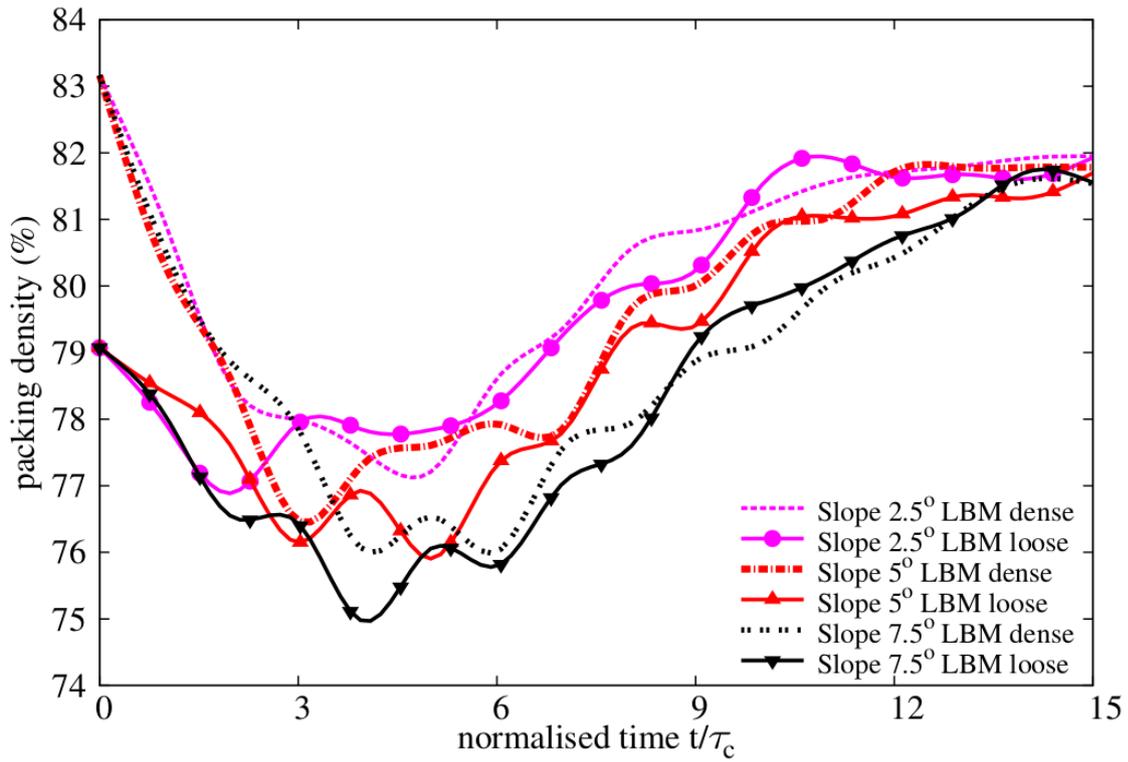

(a) Evolution of packing density.

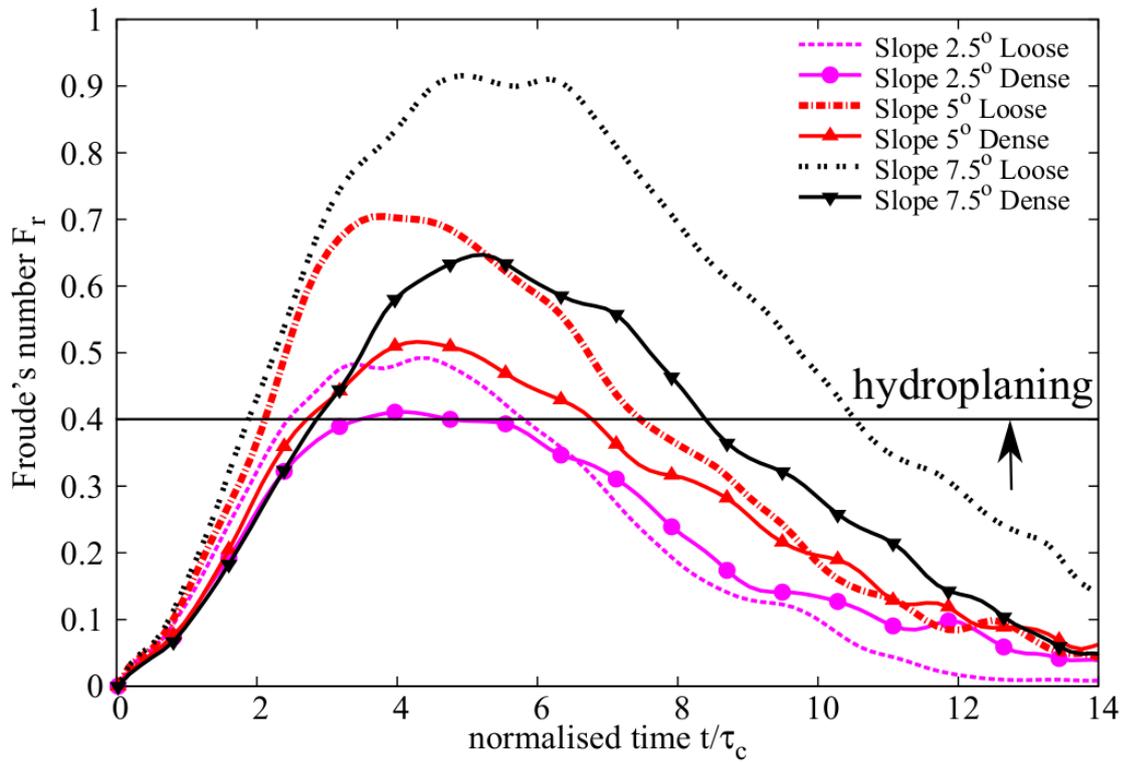

(b) Evolution of Froude's number.

Figure 20 Evolution of packing density and Froude's number with time for different slope angles

### 6.2.3 Turbulence and drag

Closer observation of the surface of the flow front in the dense column (Figure 7) and the loose column (Figure 18) cases shows that the number of turbulent vortices formed on the surface of the flow is significantly higher in the dense column cases than in the loose column cases. This can be attributed to the triangular profiles of the flowing mass of the dense column cases, in contrast to the more parabolic profiles observed in the loose granular column cases. This increase in the interaction between the flowing granular mass and the neighbouring fluid results in an increased momentum transfer and thus formation of turbulent vortices. Figure 19a shows higher density of streamlines at the top surface for the dense granular flow, which indicates increased drag resistance. This in turn reduces the velocity and the final runout distance. Viscous drag effect is found to reduce the acceleration up to about 30% in underwater landslides [23]. In contrast the loose granular flowing mass experiences less drag, thereby reaching higher kinetic energy and longer runout distances. These mechanisms explain the loose granular masses in fluid spreading almost twice as long as the dense granular masses, as also observed in experiments [24].

### 6.3 Settlement phase

The evolution of the overall packing density for the dry and loose cases for different slope angles was shown in Figure 20a. In both cases, initially the packing density decreases with time during the runout initiation process (t= 0 to $3\tau_c$). As the body starts to slow down, the density then increases. At the end of the flow, both the dense and the loose cases reach a similar packing density, indicating that the initial density effect is erased and the granular mass reaches the critical state [25]. As the granular mass settles, the eddies which developed during the horizontal acceleration phase starts to depart away from the granular surface. Force chains start to reappear at this phase, as individual particles re-establish contacts, and the shear resistance of the granular mass increases.

# Summary and Conclusions

Two-dimensional LB-DEM simulations were performed to understand the effect of initial volume fraction on the behaviour of granular column collapse on slopes submerged in fluid. The cases were run with a granular material with relatively low permeability by adopting a hydraulic radius of 0.9R. By doing so, it was possible to examine the effect of coupled interaction of pore pressure and interparticle force during different stages of granular column collapse.

LBM-DEM simulation results show that a dense granular column behavior was initially affected by the development of negative pore pressure, which resulted in increased interparticle shear resistance. This caused the delay in initiating the mass flow and reduced the kinetic energy to move laterally. During the runout phase, the flow front had an acute angle creating more vortices and drag forces along the top of the granular mass, which resisted the movement of the granular mass. This in turn resulted in shorter runout distances when compared to the dry granular column collapse cases.

In the loose granular column collapse cases, positive pore pressures were observed inside the granular body as it sheared during the initiation phase. This allowed the body to initiate the movement faster than the dense column cases. During the runout phase, water was entrained at the flow front due to its particular parabolic shape and the packing density decreased, causing reduction in the interparticle forces at the flow front and creating some lubrication-like effect. The friction between the moving body and the bottom surface was also reduced to zero as there was no effective stress acting on the sliding plane (i.e. hydroplaning). The combined effect of positive pore pressure development, water entrainment in the front body and the hydroplaning resulted in longer runout distance compared to the dry granular column collapse cases that had the same loose packing density. An opposite behaviour was observed in the dense granular column cases, which shows the importance of initial packing density in the development of runout mechanisms.

In both dense and loose granular column cases, the final packing density was the same, indicating some critical state like behaviour, in which the initial packing memory is lost after large shearing of granular material.

# Acknowledgements

The authors would like to thank Professor Farhang Radjai, LMGC, University of Montpellier 2, France for stimulating discussions regarding this work. The first author would like to thank the Cambridge Commonwealth, Overseas Trust and the Shell-Cambridge-Brazil collaboration for the financial support to pursue this research.

**Conflict of interest: none**